\documentclass[draftclsnofoot, onecolumn, 12pt]{IEEEtran}
\usepackage{cite}
\usepackage{graphicx}
\usepackage{psfrag}
\usepackage{subfigure}
\usepackage{amsmath}
\usepackage{amssymb}
\usepackage{epsfig}
\usepackage{color}
\usepackage{epsf}
\usepackage{algorithmic}
\usepackage{algorithm}
\usepackage{epsfig}

\usepackage{fancyhdr}
\usepackage{setspace}

\usepackage{fancyhdr}
\usepackage{amsfonts}
\usepackage{times}
\usepackage{latexsym}
\usepackage{dsfont}
\usepackage{amssymb}
\usepackage{amsmath}
\usepackage{cite}
\usepackage{verbatim}
\usepackage{subfigure}
\newtheorem{theorem}{Theorem}

\newtheorem{corollary}[theorem]{Corollary}

\newtheorem{lemma}[theorem]{Lemma}

\def\bb0{{\mathbb{0}}}

\def\ba{{\mathbf{a}}}
\def\bb{{\mathbf{b}}}

\def\bd{{\mathbf{d}}}

\def\bff{{\mathbf{f}}}

\def\bs{{\mathbf{s}}}

\def\bv{{\mathbf{v}}}
\def\bw{{\mathbf{w}}}
\def\bx{{\mathbf{x}}}
\def\by{{\mathbf{y}}}

\def\b0{{\mathbf{0}}}

\def\bA{{\mathbf{A}}}

\def\bF{{\mathbf{F}}}
\def\bG{{\mathbf{G}}}
\def\bH{{\mathbf{H}}}
\def\bI{{\mathbf{I}}}

\def\bU{{\mathbf{U}}}
\def\bV{{\mathbf{V}}}

\def\bX{{\mathbf{X}}}
\def\bY{{\mathbf{Y}}}

\def\sf0{{\mathsf{0}}}

\begin{document}

\title{Interference Alignment with Analog Channel State Feedback
\thanks{The authors are with the Wireless Networking and Communications Group, Department of Electrical and Computer Engineering, The University of Texas at Austin, Austin, TX 78712 USA (e-mail: \{omarayach, rheath\}@mail.utexas.edu)}\thanks{This work was supported by the Office of Naval Research (ONR) under grant N000141010337.}}

\author{Omar~El~Ayach and Robert~W.~Heath,~Jr.}

\maketitle


\begin{abstract}

Interference alignment (IA) is a multiplexing gain optimal transmission strategy for the interference channel. While the achieved sum rate with IA is much higher than previously thought possible, the improvement often comes at the cost of requiring network channel state information at the transmitters. This can be achieved by explicit feedback, a flexible yet potentially costly approach that incurs large overhead. In this paper we propose analog feedback as an alternative to limited feedback or reciprocity based alignment. We show that the full multiplexing gain observed with perfect channel knowledge is preserved by analog feedback and that the mean loss in sum rate is bounded by a constant when signal-to-noise ratio is comparable in both forward and feedback channels. When signal-to-noise ratios are not quite symmetric, a fraction of the multiplexing gain is achieved. We consider the overhead of training and feedback and use this framework to optimize the system's effective throughput. We present simulation results to demonstrate the performance of IA with analog feedback, verify our theoretical analysis, and extend our conclusions on optimal training and feedback length.

\end{abstract}


\section{Introduction}

The interference channel is an information theoretic concept that models wireless networks in which several transmitters simultaneously communicate data to their paired receivers. The traditional approach for communication in such channels was orthogonalization, where resources are split among the various transceiver pairs. Unfortunately, this results in decaying per-user data rates as networks grow. Recent work in information theory, however, has shown that the interference channel can support sum rates that, under certain assumptions, scale linearly with the number of users at high signal-to-noise ratios (SNR). This linear sum rate scaling can be achieved by a transmission strategy known as interference alignment (IA)~\cite{Cadambe2008}. 

Interference alignment is a linear precoding technique that attempts to align interfering signals in time, frequency, or space. In multiple-input multiple-output (MIMO) networks, IA utilizes the spatial dimension offered by multiple antennas for alignment. By aligning interference at all receivers, IA reduces the dimension of interference allowing users to cancel interference via linear techniques and decode their desired signals interference free. In a single-input single-output (SISO) system with infinitely many channel extensions, IA allows users to communicate at approximately half their interference free rate~\cite{Cadambe2008}. IA has also been shown to achieve maximum multiplexing gain in a class of extended MIMO channels~\cite{GuoJafar}. While such a general result has not been proven for the constant MIMO channel, antennas have been shown to provide a practical source of dimensionality~\cite{Peters2010, Yetis2009, Ayach2009}. While various constant MIMO IA solutions have been proposed, all solutions assume some form of channel state information (CSI) at the transmitter obtained either implicitly, through channel reciprocity~\cite{Gomadam2008, Berry-BidirectionalIA}, or explicitly through interference pricing \cite{MMSE-IA} or CSI feedback~\cite{Cadambe2008, Choi2009, Peters2010, Tresch, santamaria-max-sum-rate, dimakis}. Reciprocity, however, does not hold in frequency duplexed systems. Moreover, with time duplexing, reciprocity requires tight calibration of RF devices. Calibration must either be accounted for in the RF hardware~\cite{mimo_hardware_recip}, or done dynamically using a series of forward and reverse transmissions as in~\cite{guillaud_recip} which again incurs overhead. Since reciprocity may not hold, methods to transfer CSI back to the transmitters are important for realizing the sum capacities promised by IA.

CSI feedback potentially incurs large overhead which reduces the effective data rates achieved. Therefore, low overhead feedback strategies that preserve IA's sum rate performance must be used to satisfy the CSI requirement. Unfortunately, work on managing the overhead of IA, however, is rather limited. In \cite{Berry-BidirectionalIA}, the overhead of the proposed reciprocity-based algorithm is numerically analyzed. Others have proposed network design strategies such as partitioning to reduce overhead~\cite{Peters2010a}. Network partitioning, however, only reduces the number of channels that are to be fed back and makes no attempt at improving the underlying feedback strategy itself. A well established method to improve feedback and further reduce overhead is to employ channel state quantization. For example,~\cite{Thukral2009} uses Grassmannian codebooks to directly quantize and feedback the wideband channel coefficients in single antenna systems which employ IA by coding over frequency. This limited feedback approach was later extended in~\cite{Krishnamachari2009} to systems with multiple antennas at the transmitters, receivers, or both. In both~\cite{Thukral2009} and~\cite{Krishnamachari2009}, multiplexing gain is preserved by scaling the number of feedback bits with SNR. The need for such scaling is a characteristic of digital feedback and is indeed in line with earlier results for other multiuser channels~\cite{Jindal2006}. The complexity of quantized feedback, however, increases with codebook size, which is challenging since large Grassmannian codebooks are hard to design and encode.

In this paper we propose using IA with analog feedback~\cite{Marzetta2006, Samardzija2006, caire710multiuser, shanechi-comparison-of-practical-feedback}. Instead of quantizing the channel state information, analog feedback directly transmits the channel matrix elements as uncoded quadrature and amplitude modulated symbols. In the proposed strategy, with knowledge of the forward channels, receivers train the reverse links and feedback the forward channel matrices using analog feedback.
We show that, under mild assumptions on feedback quality, performing IA on the channel estimates obtained via analog feedback incurs no loss in multiplexing gain~\cite{Ayach2010}. Specifically, using a cooperative analog feedback strategy, we show that when the SNR on both forward and reverse channels is comparable, the loss in sum rate achieved by a linear zero-forcing receiver is upper bounded by a constant which implies the preservation of multiplexing gain. We extend the performance analysis to demonstrate that other analog feedback strategies, which assume no cooperation, perform similarly. Moreover, we show that when such symmetry of SNR is not possible, the system still achieves a fraction of its degrees of freedom. Using the derived bounds on loss in sum rate due to imperfect CSI, we consider the system's effective throughput, accounting for the overhead incurred by training and feedback. We use our constructed framework to optimize training and feedback lengths to find the best operating point on the trade-off curve between overhead and sum rate. 

This paper extends the analog feedback framework that was originally proposed for the MISO broadcast channel~\cite{Marzetta2006} to the MIMO interference channel, and characterizes its performance when IA is used as a transmission strategy. We show that analog feedback is a viable alternative to the quantization based schemes presented in~\cite{Thukral2009, Krishnamachari2009}. As an advantage, analog feedback does not suffer from the same increasing complexity at high SNR as digital feedback, and the required feedback scaling will come naturally in most wireless ad hoc networks. Moreover, analog feedback remains optimal when the number of feedback symbols equals the number of feedback channel uses~\cite{caire710multiuser}. We also study the effect of feedback and training overhead on IA's performance, a concept which, with the exception of \cite{Peters2010a, Berry-BidirectionalIA}, has been mostly neglected. 


Throughout this paper we use the following notation: $\bA$ is a matrix; $\ba$ is a vector; $a$ is a scalar; $\bA^*$ and $\ba^*$ denote the conjugate transpose of $\bA$ and $\ba$ respectively; $\|\bA\|_F$ is the Frobenius norm of $\bA$; $\|\ba\|_p$ is the $p$-norm of $\ba$; $\bI_N$ is the $N \times N$ identity matrix; $\mathbb{C}^N$ is the $N$-dimensional complex space; $\mathcal{CN}(\ba,\bA)$ is a complex Gaussian random vector with mean $\ba$ and covariance matrix $\bA$; $(a_1, \hdots, a_k)$ is an ordered set. Other notation is defined when needed.

\section{System Model and Background} \label{sec:background}

In this section we introduce the MIMO interference channel under consideration. We review the concept of IA with perfect channel state information, with a focus on the equations and properties that will be used in our analysis of analog feedback in later sections.

\subsection{Interference Channel Model} \label{sec:SigModel}

\begin{figure}
  \centering
	\includegraphics[width=6in]{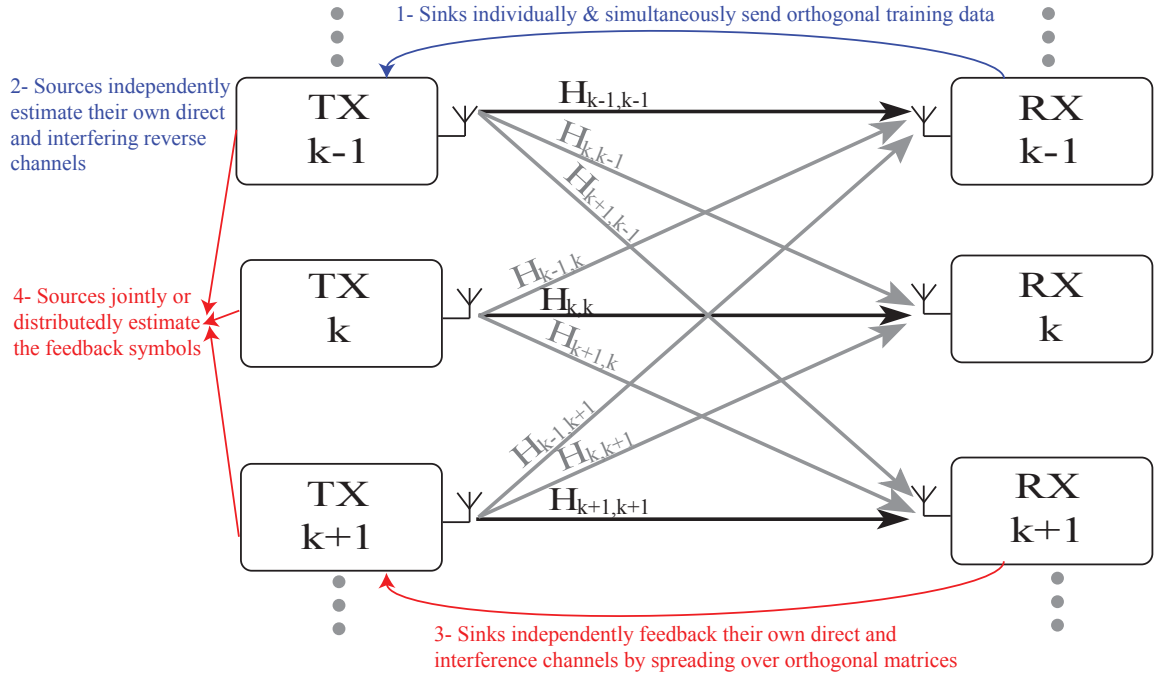}
	\caption{K-User MIMO interference channel with analog feedback.}
	\label{fig:sys_model}
\end{figure}

Consider the narrowband MIMO interference channel shown in Fig. \ref{fig:sys_model}. In this channel each of the $K$ source nodes $i$ communicates with its sink node $i$ and interferes with all other sink nodes, $\ell \neq i$. We refrain from using the term transmitter-receiver since all nodes will be involved in both transmission and reception in either payload transmission or feedback intervals. For simplicity of exposition, we consider the case of the homogeneous network where each source and sink is equipped with $N_t$ and $N_r$ antennas respectively. Thus, source node $i$ transmits $d_i\leq min(N_t,N_r)$ independent spatial streams to its corresponding sink. The results can be readily generalized to a different number of antennas at each node, provided that IA remains feasible \cite{Yetis2009}.


We consider a block fading channel model in which channels are drawn independently across all users and antennas and remain fixed for the interval of interest. We neglect large scale fading effects which can be accounted for at the expense of more involved exposition in Section \ref{sec:analog}. We also assume perfect time and frequency synchronization when expressing received baseband signals. Under our assumptions, the received signal at sink  node $i$ can be written as
\begin{equation}
\mathbf{y}_{i} = \sqrt{\frac{P}{d_i}}\mathbf{H}_{i,i}\mathbf{F}_{i}\mathbf{s}_{i} + \sum_{\ell \neq i} \sqrt{\frac{P}{d_\ell}}\mathbf{H}_{i,\ell}\mathbf{F}_{\ell}\mathbf{s}_{\ell} + \mathbf{v}_{i}, \nonumber
\label{eqn:narrowsig_model}
\end{equation}
where $\mathbf{y}_{i}$ is the $N_r \times 1$ received signal vector, $\mathbf{H}_{i,\ell}$ is the $N_r \times N_t$ channel matrix from source $\ell$ to sink $i$ with i.i.d $\mathcal{CN}(0,1)$ elements, $\mathbf{F}_{i}=[\bff_i^1,\cdots,\bff_i^{d_i}]$ is node $i$'s $N_t \times d_i$ unitary precoding matrix, $\bs_{i}$ is the $d_i \times 1$ transmitted symbol vector at node $i$ such that $\mathbb{E}\left[\|\bs_i\|^2\right]=d_i$, and $\mathbf{v}_{i}$ is a complex vector of i.i.d complex Gaussian noise with covariance matrix $\sigma^2\mathbf{I_{N_r}}$. We assume equal power allocation since the gain observed from water-filling is negligible at high SNR~\cite{martinian-waterfilling}.

We place no reciprocity assumption on the forward and reverse channels, as in a frequency division duplexed system, for example. On this channel, the received signal at \emph{source node} $i$ is
\begin{equation}
\overleftarrow{\mathbf{y}}_{i}=\sqrt{\frac{P_f}{N_r}}\bG_{i,i}\overleftarrow{\bx}_{i} + \sum_{\ell \neq i} \sqrt{\frac{P_f}{N_r}}\bG_{\ell,i}\overleftarrow{\bx}_{\ell} + \mathbf{\nu}_{i},
\label{eqn:reverse_model}
\end{equation}
where $P_{f}$ is the transmit power used to transmit pilot and feedback symbols, $\bG_{\ell,i}$ is the $N_t \times N_r$ reverse channel between sink node $\ell$ and source node $i$ with i.i.d $\mathcal{CN}(0,1)$ elements
, $\overleftarrow{\bx}_i$ is the symbol vector with unit norm elements sent by sink $i$, and $\mathbf{\nu}_{i}$ is a complex vector of i.i.d circularly symmetric white Gaussian noise with covariance matrix $\sigma^2 \mathbf{I_{N_t}}$.

\subsection{Interference Alignment} \label{sec:IA}

IA for the MIMO interference channel is a linear precoding technique that by potentially coding over infinite channel extensions achieves the channel's degrees of freedom defined as $\substack{\lim \\P\rightarrow\infty} \frac{R_{sum}}{\log_2 P}$. This result originally assumed that the magnitude of continuously distributed i.i.d channel coefficients is bounded away from zero and infinity~\cite{GuoJafar} to avoid the degenerate cases of equal coefficients or channels equal to zero or infinity. For constant MIMO channels with Rayleigh fading, degenerate cases happen with probability zero and IA can improve the achieved sum rate as shown in~\cite{Peters2010, Gomadam2008,Ayach2009}. In constant MIMO channels, IA computes the transmit precoders $\bF_i$ to align interference at all receivers in a strict subspace of the received signal space, thus leaving interference free dimensions for the desired signal. While IA is only one of the many precoding strategies for the interference channel~\cite{Rose_Greedy, Peters2010, Berry-BidirectionalIA}, some of which marginally outperform it at low SNR~\cite{Peters2010}, a main advantage of IA is that it is analytically tractable. Its complete interference suppression properties make it especially amenable to the study of performance with feedback and imperfect CSI. 


While IA can be used with any receiver design, to simplify exposition we consider a per-stream zero-forcing receiver in which sink node $i$ projects its received signal on to the columns, $\bw_i^m$, and treats residual interference as noise. Simulations in Section \ref{sec:sims} indicate that the same performance can be expected from a joint signal decoder which again treats interference as noise.

Writing the per stream input-output relation at the output of $\bw_i^m$ gives
\begin{equation}
(\bw_i^m)^*\by_i=(\bw_i^m)^*\sqrt{\frac{P}{d_i}}\mathbf{H}_{i,i}\bff_{i}^{m} s_{i}^{m}+\sum\limits_{\ell\neq m}(\bw_i^m)^*\sqrt{\frac{P}{d_i}}\mathbf{H}_{i,i}\bff_{i}^{\ell} s_{i}^{\ell} +\sum\limits_{k\neq i}\sum\limits_{\ell=1}^{d_k}(\bw_i^m)^*\sqrt{\frac{P}{d_k}}\mathbf{H}_{i,k}\bff_{k}^{\ell} s_{k}^{\ell}+(\bw_i^m)^*\bv_i, \nonumber
\label{eqn:recv_proj}
\end{equation}
for $m\in\left\{1,\ldots,d_i\right\}$ and $i \in \left\{1,\ldots,K\right\}$, where $\|\bw_i^m\|^2=1$. At the output of these linear receivers $\bw_i^m$, the conditions for perfect interference alignment can be restated as \cite{Gomadam2008}
\begin{eqnarray}
(\bw_i^m)^*\mathbf{H}_{i,k}\bff_{k}^{\ell}=0, &\quad& \forall (k,\ell) \neq (i,m) \label{eqn:conditions1}\\
\left|(\bw_i^m)^*\mathbf{H}_{i,i}\bff_{i}^{m}\right|\geq c >0, &\quad& \forall i,m \label{eqn:conditions2}
\label{eqn:conditions}
\end{eqnarray}
where IA is guaranteed by the first condition, and the second is satisfied with high probability.

The sum rate achieved by such a linear zero-forcing receiver, assuming Gaussian input signals and treating leakage interference as noise, is
\begin{equation}
R_{sum} = \sum\limits_{i=1}^{K}\sum\limits_{m=1}^{d_i}
\log_2\left(1+\frac{\frac{P}{d_i}\left|(\bw_i^m)^*\mathbf{H}_{i,i}\bff_{i}^{m}\right|^2}{\mathcal{I}_{i,m}+\sigma^2}\right),
\label{eqn:sumrate}
\end{equation}
where $\mathcal{I}_{i,m}$ is the total inter-stream and inter-user interference given by \cite{Jindal2006, caire710multiuser}
\begin{eqnarray}
\mathcal{I}_{i,m} = \sum\limits_{(k,\ell)\neq (i,m)}\frac{P}{d_k} \left|(\bw_i^m)^*\mathbf{H}_{i,k}\bff_{k}^{\ell}\right|^2. \nonumber
\end{eqnarray}
In the presence of perfect channel knowledge, and for an achievable degree of freedom vector $\bd=[d_1, d_2, \ldots, d_K]$, equations (\ref{eqn:conditions1}) and (\ref{eqn:conditions2}), are satisfied and thus $\mathcal{I}_{i,m}=0$. This gives
\begin{align}
\begin{split}
\lim_{P\rightarrow\infty} \frac{R_{sum}}{\log_2 P}  = \lim_{P\rightarrow\infty} \frac{\sum\limits_{i,m}\log_2\left(1+\frac{\frac{P}{d_i}\left|(\bw_i^m)^*\mathbf{H}_{i,i}\bff_{i}^{m}\right|^2}{\sigma^2}\right)}{\log_2 P} = \sum\limits_{i=1}^{K} d_i\leq\mathrm{min}(N_t,N_r)\frac{R}{R+1}K,\nonumber
\end{split}
\end{align}
in the case where $R=\frac{\max(N_t,N_r)}{\min(N_t,N_r)}$ is an integer such that $K>R$ \cite{GuoJafar}. The full characterization of the extended MIMO interference channel's degrees of freedom is provided in \cite{GuoJafar}.

It is not immediately clear, however, if the same sum rate scaling behavior can be expected from a network with only imperfect knowledge of the channel derived from analog feedback. Results on single user MIMO generally prove an acceptable constant rate loss due to imperfect CSI~\cite{love-heath-limited-feedback-unitary}. In most multi-user scenarios, however, the cost of imperfect CSI may be much higher, potentially resulting in the loss of the channel's multiplexing gain~\cite{Jindal2006} which saturates performance at high SNR~\cite{makouei-simple-sinr-characterization}. In Section \ref{sec:analogIA}, we show that such performance can be expected from a realistic IA system with analog feedback, provided that the quality of channel knowledge scales sufficiently with transmit power, or effectively the forward channel's SNR. The result is based on the necessary accuracy of CSI which must scale with SNR as shown in~\cite{Caire_shamai}. This is similar to the results presented in~\cite{Thukral2009} and \cite{Krishnamachari2009} for IA with digital feedback where CSI quality is scaled by controlling codebook size and feedback bits. 


\section{Interference Alignment with Analog Feedback} \label{sec:analogIA}

In this section we propose a feedback and transmission strategy based on analog feedback and naive IA, which uses the estimated channels as if they were the true propagation channels.

\subsection{Analog Feedback} \label{sec:analog}

To feedback the forward channel matrices $\bH_{i,\ell}$ reliably across the feedback channel given in (\ref{eqn:reverse_model}), we propose dividing the feedback stage into two main phases~\cite{Marzetta2006}. The details of the analog feedback strategy considered are summarized in Fig. \ref{fig:sys_model}. First, each source independently learns its reverse channels. Second, the forward channels are fed back and estimated. We neglect forward channel training and estimation and, thus, assume they have been estimated perfectly. This is since imperfect CSI at the receiver adversely affects all feedback schemes and is not exclusive to analog feedback. In fact, the analysis of forward channel estimation error parallels the analysis in Section \ref{sec:reverse_pilot} and will simply add an extra error term that decays with power as needed in Section \ref{sec:mux_proof}. Therefore, similar results can be readily shown. 

\subsubsection{Reverse Link Training} \label{sec:reverse_pilot}

To learn all reverse links, the $K$ sink nodes must transmit known pilot symbols over a period $\tau_p \geq KN_r$. Similar to the analysis done in~\cite{Marzetta2006}, we let each sink independently and simultaneously transmit an $N_r \times \tau_p$ matrix of pilots $\mathbf{\Phi}_i$ such that $\mathbf{\Phi}_i\mathbf{\Phi}_k^*=\delta_{ik}\bI_{N_r}$ shown to be optimal in~\cite{marzetta1999blast}. This training phase only requires synchronization. 

Let $\overleftarrow{\bY}_i=\left[\overleftarrow{\by}_i[1]\ \ldots\ \overleftarrow{\by}_i[\tau_p]\right]$ be the $N_t \times \tau_p$ received training matrix at source node $i$. Then the received training is
\begin{equation}
\overleftarrow{\bY}_i=\sqrt{\frac{\tau_p P_{f}}{N_r}}\sum\limits_{k=1}^{K}\bG_{k,i}\mathbf{\Phi}_k+\bV_i, \quad \forall i \nonumber
\label{eqn:training}
\end{equation}
where $\bV_i$ is an $N_t \times \tau_p$ matrix of i.i.d $\mathcal{CN}(0,\sigma^2)$ noise elements. Using its received training, each source $i$ locally computes MMSE estimates of its channels given by
\begin{equation}
\widehat{\bG}_{k,i}=\frac{\sqrt{\frac{\tau_p P_{f}}{N_r}}}{\sigma^2+\frac{\tau_p P_{f}}{N_r}}\overleftarrow{\bY}_i\mathbf{\Phi}_k^* \quad \forall k.
\label{eqn:reverse_mmse}
\end{equation}
Since $\widehat{\bG}_{k,i}$ are MMSE estimates of Gaussian variables corrupted by Gaussian noise, this results in $\widehat{\bG}_{k,i} \sim  \mathcal{CN}\left(0,\frac{\frac{\tau_p P_{f}}{N_r}}{\sigma^2+\frac{\tau_p P_{f}}{N_r}}\right)$ and $\widetilde{\bG}_{k,i}=\bG_{k,i}-\widehat{\bG}_{k,i}$ with i.i.d $\mathcal{CN}\left(0,\frac{\sigma^2}{\sigma^2+\frac{\tau_p P_{f}}{N_r}}\right)$ elements. 

\subsubsection{Analog CSI Feedback} \label{sec:CSI_fb}

After reverse link training, each sink node $i$ independently sends its unquantized uncoded estimates of $\bH_{i,k}\ \forall\ k$ over a period $\tau_c$. To have the sink nodes feedback their CSI simultaneously and orthogonally, each sink post multiplies its $N_r \times KN_t$ feedback matrix $\left[\bH_{i,1} \ldots \bH_{i,K}\right]$ with a $KN_t \times \tau_c$ matrix $\mathbf{\Psi}_i$ such that $\mathbf{\Psi}_i\mathbf{\Psi}_k^*=\delta_{i,k}\bI_{KN_t}$~\cite{Marzetta2006}. This is a general orthogonal structure that can capture the case of orthogonality in time and requires $\tau_c \geq K^2N_t$. The transmitted $N_r \times \tau_c$ feedback matrix $\overleftarrow{\bX}_i$ from sink $i$ can be written as
\begin{equation}
\overleftarrow{\bX}_i=\sqrt{\frac{\tau_c P_{f}}{KN_tN_r}}\left[\bH_{i,1} \ldots \bH_{i,K}\right]\mathbf{\Psi}_i. \nonumber
\end{equation}
The concatenated  $KN_t \times \tau_c$ matrix of received feedback is then given by
\begin{equation}
\overleftarrow{\bY}_c=\sqrt{\frac{\tau_c P_{f}}{KN_tN_r}}\sum\limits_{i=1}^{K}\left[\begin{array}{c} \bG_{i,1} \\ \vdots \\ \bG_{i,K}\end{array}\right]\left[\bH_{i,1} \ldots \bH_{i,K}\right]\mathbf{\Psi}_i + \bV \nonumber
\end{equation}
where $\bV$ is now a $KN_t \times \tau_c$ noise matrix.


To estimate the forward channels $\bH_{i,k}$, the source nodes first isolate the training from sink node $i$ by post multiplying their received training by $\mathbf{\Psi}_i^*$ which gives
\begin{equation}
\overleftarrow{\bY}_c\mathbf{\Psi}_i^*=\sqrt{\frac{\tau_c P_{f}}{KN_tN_r}}\underbrace{\left[\begin{array}{c} \bG_{i,1} \\ \vdots \\ \bG_{i,K}\end{array}\right]}_{\bG_i}\underbrace{\left[\bH_{i,1} \ldots \bH_{i,K}\right]}_{\bH_i} + \bV\mathbf{\Psi}_i^*.
\label{eqn:Ypsi}
\end{equation}
To simplify the analysis in Section \ref{sec:mux_proof}, we assume that the complete received feedback matrix $\overleftarrow{\bY}_c$ is shared among sources. Then, assuming $KN_t \geq N_r$, sources compute a common least squares estimate $\widehat{\bH}_i$ of $\bH_i$ given by
\begin{eqnarray}
\widehat{\bH}_i &=& \sqrt{\frac{KN_tN_r}{\tau_c P_{f}}}\left(\widehat{\bG}_i^* \widehat{\bG}_i\right)^{-1} \widehat{\bG}_i^*\overleftarrow{\bY}_c\mathbf{\Psi}_i^* \nonumber\\
&=& \underbrace{\bH_i}_\textrm{Real Channel} + \underbrace{\widetilde{\bH}_i}_\textrm{Error}, \nonumber
\end{eqnarray}
where $\widehat{\bG}_i$ is the MMSE estimate of $\bG_i$. Such node cooperation is only realistic in certain cases such as IA for cellular systems for example. At the end of this section, we provide alternative non-cooperative approaches that we show in Section \ref{sec:sims} perform close to this special case.

The error in the estimates of $\bH_{i}$ can then be written as
\begin{equation}
\widetilde{\bH}_i = \left(\widehat{\bG}_i^* \widehat{\bG}_i\right)^{-1} \widehat{\bG}_i^*\left(\sqrt{\frac{KN_tN_r}{\tau_c P_{f}}}\bV + \widetilde{\bG}_i\bH_i\right). \nonumber
\end{equation}
which makes it clear that the error in the estimate consists of two error terms: the first due to noisy feedback and the second due to a noisy estimate of the feedback channel. To quantify the effect of the error on the achieved sum rate, we derive the variance of the error term introduced by analog feedback. Recall that the elements of $\bH_{i,k}$ are $\mathcal{CN}(0,1)$, those of $\bV$ are $\mathcal{CN}(0,\sigma^2)$, and those of $\widetilde{\bG}_i$ are $\mathcal{CN}(0,\frac{\sigma^2}{\sigma^2+\frac{\tau_p P_{f}}{N_r}})$. As a result, the error term $\widetilde{\bG}_i\bH_i$ due to the reverse channel estimation has independent elements with a variance of $\frac{N_r \sigma^2}{\sigma^2+\frac{\tau_p P_{f}}{N_r}}$. Similarly to~\cite{Marzetta2006} we see that the covariance of each column of $\widetilde{\bH}_i$ denoted $\widetilde{\bH}_i^{(\ell)}$, conditioned on $\widehat{\bG}_i$ is 
$$
\textrm{Cov}(\widetilde{\bH}_i^{(\ell)} | \widehat{\bG}_i)=\left(\frac{KN_tN_r\sigma^2}{\tau_c P_f}+\frac{N_r\sigma^2}{\sigma^2+\tau_p \frac{P_f}{N_r}}\right)\left(\widehat{\bG}_i^*\widehat{\bG}_i\right)^{-1}.
$$
Since the elements of the MMSE estimate $\widehat{\bG}_i$ are Gaussian and uncorrelated, the diagonal elements of $\left(\widehat{\bG}_i^*\widehat{\bG}_i\right)^{-1}$ are reciprocals of scaled chi-squared random variables with $2(KN_t-N_r+1)$ degrees of freedom. As a result, the mean square error, $\sigma_{f}^2$, in the elements of $\widehat{\bH}_{i,k}$ is given by
\begin{equation}
\sigma_{f}^2= \frac{\sigma^2}{(KN_t-N_r)P_{f}}\left(\frac{N_r^2}{\tau_p}+\frac{KN_tN_r}{\tau_c}\left(1+\frac{N_r\sigma^2}{\tau_p P_{f}}\right)\right).
\label{eqn:MSE}
\end{equation}
At high SNR this gives
\begin{equation}
\sigma_{f}^2\approx\frac{\sigma^2\left(\frac{N_r^2}{\tau_p}+\frac{KN_tN_r}{\tau_c}\right)}{(KN_t-N_r) P_{f}}.
\label{eqn:MSE_High_SNR}
\end{equation}


Having computed feedback error, we return to the assumption on node cooperation. Cooperation simplifies Section III-B by making a common channel estimate known to all users. Such cooperation may not always be possible. We present two alternative practical strategies, outline the changes they incur to the feedback process, and discuss their shortcomings:
\begin{itemize}
\item Centralized processor: In this scheme, reverse link training, estimation and feedback transmission remain unchanged. In this strategy however, the $KN_t\times \tau_c$ matrix  $\overleftarrow{\bY}_c$  is not shared, and instead one of the nodes calculates an estimate of the feedback information based on its locally observed $N_t\times \tau_c$ matrix of feedback symbols (i.e. only its own rows of $\overleftarrow{\bY}_c$). This can be done as long as $N_t\geq N_r$\footnote{The restriction $N_t\geq N_r$ can be relaxed at the expense of a longer $\tau_c$ by assigning each user several spreading matrices $\mathbf{\Psi}$.} and results in $\sigma_{f}^2\approx\frac{\sigma^2}{(N_t-N_r)P_f}\left(\frac{N_r^2}{\tau_p}+\frac{KN_tN_r}{\tau_c}\right)$. This source then calculates the IA solution required and feeds it forward, again using ``analog feedforward'', to all other sources~\cite{chae2008coordinated}.
\item Distributed processing: In this strategy reverse link training, estimation and feedback transmission again remain unchanged. In this scheme, however, \emph{each} source locally calculates its own estimate of all the channels being fed back, using the $N_t\times \tau_c$ symbols it independently receives. Therefore, each node will obtain a different perturbed estimate of the forward channels with the same $\sigma_f^2$ as in the centralized processor case. Each node then locally calculates a complete set of perturbed IA precoders based on its channel estimates.
\end{itemize}
We expand on the performance of these two alternative practical strategies in Section V.

\subsection{Multiplexing Gain with Analog Feedback} \label{sec:mux_proof}

To characterize the performance of IA with analog feedback, we examine the rate loss~\cite{Jindal2006} incurred by naive IA where channel estimates are used to calculate the columns of the precoders, $\widehat{\bff}_i^m \ \forall i,m$ and combiners $\widehat{\bw}_i^m \ \forall i,m$. Therefore, such transmit and receive vectors satisfy 
\begin{eqnarray}
(\widehat{\bw}_i^m)^*\widehat{\bH}_{i,k}\widehat{\bff}_{k}^{\ell}=0, &\quad& \forall (k,\ell)\neq (i,m)  \label{eqn:conditions1hat}\\
\left|(\widehat{\bw}_i^m)^*\widehat{\bH}_{i,i}\widehat{\bff}_{i}^{m}\right|\geq c >0, &\quad& \forall i,m,\label{eqn:conditions2hat}
\end{eqnarray}
and can be found for a feasible degree of freedom vector $\bd=\left[d_1, \hdots, d_K\right]$. As stated earlier, sinks need not use such per-stream combiners, and can employ a number of other receiver designs based on their channel knowledge. In Section \ref{sec:sims} we show the performance of a joint decoder which whitens interference using its covariance~\cite{Blum2003}. Such receiver CSI can be acquired blindly~\cite{honigblind} or via additional short training or silent phases. For the sake of the following rate loss analysis, we abstract the calculation of receiver CSI and assume that sinks have learned the combiners $\widehat{\bw}_i^m$.

The mean loss in sum-rate is then defined as $\Delta R_{sum}\triangleq\mathbb{E}_\bH R_{sum}-\mathbb{E}_\bH \widehat{R}_{sum}$,
where $\mathbb{E}_\bH R_{sum}$ is the average sum rate from IA with perfect CSI, with instantaneous rate given in (\ref{eqn:sumrate}), and $\mathbb{E}_{\bH}\widehat{R}_{sum}$ is the rate achieved with imperfect CSI and the vectors in (\ref{eqn:conditions1hat}),  (\ref{eqn:conditions2hat}). Before proving the main result on rate loss, we will need the following lemma.
\begin{lemma}
The desired signal powers, $\left|(\bw_i^m)^*\bH_{i,i}\bff_{i}^{m}\right|^2$ and $\left|(\widehat{\bw}_i^m)^*\bH_{i,i}\widehat{\bff}_{i}^{m}\right|^2$, resulting from IA with perfect or imperfect CSI respectively are identically and exponentially distributed. 
\label{lemma:identical}
\end{lemma}
\begin{IEEEproof}
See Appendix \ref{sec:proof_lemma}.
\end{IEEEproof}
\begin{theorem}
The sum rate loss experienced by IA on the $K$-user $N_r \times N_t$ interference channel with imperfect CSI obtained via the analog feedback strategy described in Section \ref{sec:analog} is upper bounded by a constant when $P_{f}$ scales with the transmit power $P$.
\label{thm:sumratescaling}
\end{theorem}
\begin{IEEEproof}
Let the $K$-user $N_r \times N_t$ interference channel use the analog feedback scheme presented to achieve a vector of multiplexing gains $\bd$. 
Using the imperfect vectors  $\widehat{\bff}_i^m$ and $\widehat{\bw}_i^m$ respectively, the input-output relationship at the output of a linear zero-forcing receiver is
\begin{equation}
(\widehat{\bw}_i^m)^*\by_i=(\widehat{\bw}_i^m)^*\sqrt{\frac{P}{d_i}}\mathbf{H}_{i,i}\widehat{\bff}_{i}^{m} s_{i}^{m}+\sum\limits_{\ell\neq m}(\widehat{\bw}_i^m)^*\sqrt{\frac{P}{d_i}}\mathbf{H}_{i,i}\widehat{\bff}_{i}^{\ell} s_{i}^{\ell}+\sum\limits_{k\neq i}\sum\limits_{\ell=1}^{d_k}(\widehat{\bw}_i^m)^*\sqrt{\frac{P}{d_k}}\mathbf{H}_{i,k}\widehat{\bff}_{k}^{\ell} s_{k}^{\ell}+(\widehat{\bw}_i^m)^*\bv_i.
\label{eqn:recv_proj_imp} \nonumber
\end{equation}
Using this received signal and the instantaneous rate expression in (\ref{eqn:sumrate}) gives the following upper bound on mean loss in sum rate:
\begin{align}
\begin{split}
\Delta R_{sum} & = \mathbb{E}_\bH\sum\limits_{i,m} \log_2\left(1+\frac{\frac{P}{d_i}\left|(\bw_i^m)^*\mathbf{H}_{i,i}\bff_{i}^{m}\right|^2}{\sigma^2}\right)
-\mathbb{E}_{\bH,\widehat{\bH}}\sum\limits_{i,m} \log_2\left(1+\frac{\frac{P}{d_i}\left|(\widehat{\bw}_i^m)^*\mathbf{H}_{i,i}\widehat{\bff}_{i}^{m} \right|^2}{\mathcal{I}_{i,m}+\sigma^2}\right) \\
& = \mathbb{E}_\bH\sum\limits_{i,m} \log_2\left(1+\frac{\frac{P}{d_i}\left|(\bw_i^m)^*\mathbf{H}_{i,i}\bff_{i}^{m}\right|^2}{\sigma^2}\right)\\
&\ \  -\mathbb{E}_{\bH,\widehat{\bH}}\sum\limits_{i,m}\log_2 \left(1+\frac{\mathcal{I}_{i,m} + \frac{P}{d_i}\left|(\widehat{\bw}_i^m)^*\mathbf{H}_{i,i}\widehat{\bff}_{i}^{m} \right|^2}{\sigma^2}\right) \\
&\  \  +\mathbb{E}_{\bH,\widehat{\bH}}\sum\limits_{i,m}\log_2\left(1+\frac{\mathcal{I}_{i,m}}{\sigma^2}\right) \\
& \stackrel{(a)}{\leq}\mathbb{E}_{\bH\widehat{\bH}}\sum\limits_{i,m}\log_2\left(1+\frac{\mathcal{I}_{i,m}}{\sigma^2}\right) 
\stackrel{(b)}{\leq}\sum\limits_{i,m}\log_2\left(1+\frac{\mathbb{E}_{\bH,\widehat{\bH}}\mathcal{I}_{i,m}}{\sigma^2}\right). \nonumber
\end{split}
\end{align}
where (a) follows from Lemma \ref{lemma:identical} which along with $\mathcal{I}_{i,m}\geq 0$ implies that $\frac{P}{d_i}\left|(\widehat{\bw}_i^m)^*\mathbf{H}_{i,i}\widehat{\bff}_{i}^{m} \right|^2+\mathcal{I}_{i,m}$ stochastically dominates $\frac{P}{d_i}\left|(\bw_i^m)^*\mathbf{H}_{i,i}\bff_{i}^{m} \right|^2$ and (b) follows from Jensen's inequality~\cite{Jindal2006}.

Since (\ref{eqn:conditions1hat}) and (\ref{eqn:conditions2hat}) are satisfied, however, the total interference term $\mathcal{I}_{i,m}$ can be simplified to include only residual interference due to the channel estimation errors $\widetilde{\bH}_{i,\ell}$. Therefore, $\Delta R_{sum}$ can be further upper bounded by noticing that
\begin{align}
\begin{split}
\mathbb{E}_{\bH,\widehat{\bH}}\left|(\widehat{\bw}_i^m)^*(\widehat{\bH}_{i,k}-\widetilde{\bH}_{i,k})\widehat{\bff}_{k}^{\ell} \right|^2 & = \mathbb{E}_{\bH,\widehat{\bH}}\left|(\widehat{\bw}_i^m)^*(\widetilde{\bH}_{i,k})\widehat{\bff}_{k}^{\ell} \right|^2 \leq \mathbb{E}_{\bH,\widehat{\bH}}\|\widetilde{\bH}_{i,k}\|_F^2, \quad \forall k,\ \forall \ell \neq m
\label{eqn:frob_bound}
\end{split}
\end{align}
which gives
\small
\begin{equation}
\Delta R_{sum} \leq \sum\limits_{i,m}\log_2\left(1 + \frac{1}{\sigma^2}\sum\limits_{\ell=1}^{K}\frac{P}{d_\ell}\left(d_\ell-\delta_{i,\ell}\right)\mathbb{E}_{\bH,\widehat{\bH}}\|\tilde{\bH}_{i,\ell}\|_F^2\right).
\label{eqn:final_loss}
\end{equation}
\normalsize
From (\ref{eqn:MSE}), however, we have $\mathbb{E}_{\bH,\widehat{\bH}}\|\tilde{\bH}_{i,\ell}\|_F^2=N_tN_r\sigma_{f}^2=\frac{c(\tau_p,\tau_c)\sigma^2}{P_{f}}$, where $c(\tau_p,\tau_c)$ is a constant, independent of $P_{f}$ at high enough SNR, given by
\begin{equation}
c(\tau_p,\tau_c)=N_tN_r\frac{\left(\frac{N_r^2}{\tau_p}+\frac{KN_tN_r(1+\epsilon)}{\tau_c}\right)}{(KN_t-N_r)}. \label{eqn:constant} 
\end{equation}
Combining (\ref{eqn:final_loss}), (\ref{eqn:constant}) and letting $P_f=\alpha^{-1}P$ gives the final upper bound on throughput loss
\small
\begin{equation}
\Delta R_{sum}(\tau_p,\tau_c) \leq \sum\limits_{i,m}\log_2\left(1+\frac{c(\tau_p,\tau_c)\sigma^2 P}{\sigma^2 P_f}\sum\limits_{\ell=1}^{K}\frac{d_\ell-\delta_{i,\ell}}{d_\ell}\right) \leq \sum\limits_{i}d_i\log_2\left(1+\alpha c(\tau_p,\tau_c)\left(K-\frac{1}{d_i}\right)\right).
\label{eqn:final_loss_alpha}
\end{equation}
\normalsize
The bound has been presented at high SNR for simplicity of exposition only; (\ref{eqn:constant}) can be adapted for any $\textrm{SNR}>0$ by using (\ref{eqn:MSE}) instead of (\ref{eqn:MSE_High_SNR}).
\end{IEEEproof}
\begin{corollary}
IA with imperfect CSI obtained via the analog feedback strategy in Section \ref{sec:analog} achieves the same average multiplexing gain as IA with perfect CSI.
\label{corr:mux_gain}
\end{corollary}
\begin{IEEEproof}
This follows immediately from the constant loss in sum rate shown in Theorem \ref{thm:sumratescaling} which becomes negligible as SNR is taken to infinity.
\end{IEEEproof}
In summary, Theorem \ref{thm:sumratescaling} states that if the SNR on the reverse and forward link are comparable, the cost of imperfect CSI is a constant. This constant is a decreasing function of $\tau_c$ and $\tau_p$ and thus, we have written $\Delta R_{sum}(\tau_p,\tau_c)$ in \eqref{eqn:final_loss_alpha} to highlight this dependence. Note that while bounding leakage interference as $\mathbb{E}_{\bH,\widehat{\bH}}\left|(\widehat{\bw}_i^m)^*(\widetilde{\bH}_{i,k})\widehat{\bff}_{k}^{\ell} \right|^2 \leq \mathbb{E}_{\bH,\widehat{\bH}}\|\widetilde{\bH}_{i,k}\|_F^2=N_tN_r \sigma_f^2$ suffices to establish a constant rate loss, it is very conservative, and increasingly loose for larger systems. If we assume that $\widetilde{\bH}_{i,k}$ has a bi-unitary invariant distribution, e.g. Gaussian, then by the same reasoning as in Appendix \ref{sec:proof_lemma} we get $\mathbb{E}_{\bH,\widehat{\bH}}\left|(\widehat{\bw}_i^m)^*(\widetilde{\bH}_{i,k})\widehat{\bff}_{k}^{\ell} \right|^2 = \sigma^2_f$ and $c(\tau_p,\tau_c)$ is replaced with $c_2(\tau_p,\tau_c)=\frac{1}{KN_t-N_r}\left(\frac{N_r^2}{\tau_p}+\frac{KN_tN_r(1+\epsilon)}{\tau_c}\right)$. Since SNRs are likely to be comparable in practice, analog feedback allows systems to overcome the problem of fast scaling complexity in digital feedback, which remains even if the actual feedback stage operates error-free close to capacity~\cite{Jindal2006, caire710multiuser}. In case such SNR symmetry is not possible, however, systems become interference limited and multiplexing gain is reduced.

\begin{theorem}
IA on a $K$-user $N_r \times N_t$ interference channel using analog feedback with $P_{f}=\alpha P^\beta$ such that $0 \leq \beta \leq1$ achieves at least a $\beta$-fraction of its original multiplexing gain.   
\label{thm:sumrate_slow_feedback}
\end{theorem}
\begin{IEEEproof}
The multiplexing gain achieved by a system using naive IA and analog feedback with $P_{f}=\alpha P^\beta$ can be written as
\begin{align}
\begin{split}
M & =\lim_{P\rightarrow \infty}\frac{\mathbb{E}_{\bH,\widehat{\bH}}[\widehat{R}_{sum}]}{\log_2 (P)}= \lim_{P\rightarrow \infty}\frac{\mathbb{E}_{\bH,\widehat{\bH}}\sum\limits_{i,m} \log_2\left(1+\frac{\frac{P}{d_i}\left|(\widehat{\bw}_i^m)^*\mathbf{H}_{i,i}\widehat{\bff}_{i}^{m} \right|^2}{\mathcal{I}_{i,m}+\sigma^2}\right)}{\log_2 (P)}\\
& =\lim_{P\rightarrow\infty} \frac{\mathbb{E}_{\bH,\tilde{\bH}}\sum\limits_{i,m} \log_2\left(\frac{P}{d_i}\left|(\widehat{\bw}_i^m)^*\mathbf{H}_{i,i}\widehat{\bff}_{i}^{m} \right|^2+ \mathcal{I}_{i,m}+\sigma^2\right) - \mathbb{E}_{\bH,\tilde{\bH}} \sum\limits_{i,m} \log_2\left(\mathcal{I}_{i,m}+\sigma^2\right)}{\log_2 (P)} \\
& \stackrel{(a)}{\geq} \frac{\mathbb{E}_{\bH,\tilde{\bH}}\sum\limits_{i,m} \log_2\left(\frac{P}{d_i}\left|(\widehat{\bw}_i^m)^*\mathbf{H}_{i,i}\widehat{\bff}_{i}^{m} \right|^2\right) - \sum\limits_{i,m} \log_2\left( \mathbb{E}_{\bH,\tilde{\bH}}\mathcal{I}_{i,m}+\sigma^2\right)}{\log_2 (P)} \\
& \stackrel{(b)}{\geq} \sum\limits_{k=1}^{K}d_k - \lim_{P\rightarrow\infty} \frac{\sum\limits_{i.m}\log_2\left(\frac{\left(d_i-1\right)P}{d_i}\mathbb{E}_{\bH,\tilde{\bH}}\|\widetilde{\bH}_{i,i}\|_F^2+(K-1)P \mathbb{E}_{\bH,\tilde{\bH}}\|\widetilde{\bH}_{i,k}\|_F^2+\sigma^2\right)}{\log_2 (P)}  \nonumber
\end{split}
\end{align}
\begin{align}
\begin{split}
& \stackrel{(c)}{=} \sum\limits_{k=1}^{K}d_k - \lim_{P\rightarrow\infty} \frac{\sum\limits_{i.m}\log_2\left(\frac{\left(d_i-1\right)P^{1-\beta}c(\tau_p, \tau_c)\sigma^2}{\alpha d_i}+\frac{(K-1)P^{1-\beta}c(\tau_p, \tau_c)\sigma^2}{\alpha}+\sigma^2\right)}{\log_2 (P)} =\beta\left(\sum\limits_{k=1}^{K}d_k\right), \nonumber
\end{split}
\end{align}
where (a) follows from disregarding interference in the first term and applying Jensen's to the second; (b) follows from (\ref{eqn:frob_bound}) and the fact that each term in the first summation has a multiplexing gain of 1. Finally, (c) is due to the scaling of residual interference with $P^{1-\beta}$.
\end{IEEEproof}

In the case where $\beta=0$, feedback power is constant, and the interference limited system achieves zero multiplexing gain. In fact, the system simulations in Section \ref{sec:sims} indicate that the sum-rate achieved is upper bounded by a constant. As for the case of $\beta=1$, Theorem \ref{thm:sumrate_slow_feedback} confirms the preservation of full multiplexing gain proven in Corollary \ref{corr:mux_gain}, but does not establish the $O(1)$ loss in sum rate of Theorem \ref{thm:sumratescaling}. As for any $0<\beta<1$, Theorem \ref{thm:sumrate_slow_feedback} shows that analog feedback and interference alignment can still achieve linear sum rate scaling even when feedback power is much smaller than transmit power. This may be the case in certain non-homogeneous networks such as cellular networks in which mobile and base station powers do not match.

\section{Degrees of Freedom with Overhead} \label{sec:overhead}

While the analysis done in Section \ref{sec:mux_proof} is a good indicator of the cost of imperfect CSI obtained through training and analog feedback, it does not directly predict the expected throughput achieved by this strategy. Namely, the analysis done thus far neglects the cost of training and feedback overhead. In this section we define the expected throughput with overhead and use it to optimize training and feedback.

\subsection{Definition of Overhead}

As shown in Section \ref{sec:mux_proof}, the performance of IA is tightly related to the mean square error in the channel estimates at the transmitter.  When operating in a time varying channel, training and feedback must be done periodically to ensure the validity of the channel estimate at the transmitter. Depending on the channel's coherence time, the overhead due to training and feedback may consume an arbitrarily large fraction of resources such as time or frequency slots, resulting in low net throughput. In this section we consider the case in which training, feedback, and data transmission are all orthogonal in time, in the same coherence time or frame $T$~\cite{Peters2010a,kobayashi2009training}. Using this model for training and feedback overhead we compute the expected effective sum rate as
\begin{equation}
R_{eff}(\tau_p,\tau_c)=\left(\frac{T-(\tau_c+\tau_p)}{T}\right)\left(\mathbb{E}_{\bH}R_{sum}-\Delta R_{sum}\right).
\label{eqn:sum_overhead}
\end{equation}
Using the rate loss expression in (\ref{eqn:final_loss_alpha}), we write the expected sum rate with overhead as
\begin{equation}
R_{eff}(\tau_p,\tau_c) \approx \left(\frac{T-(\tau_c+\tau_p)}{T}\right)\left(\mathbb{E}_{\bH}R_{sum}- \sum\limits_{i}d_i\log_2\left(1+\alpha c_2(\tau_p, \tau_c)\left(K-\frac{1}{d_i}\right)\right)\right),
\end{equation}
where we have used $c_2(\tau_p,\tau_c)$ rather than the looser $c(\tau_p,\tau_c)$. As expected, and as can be seen from (\ref{eqn:sum_overhead}), insufficient training and feedback may result in poor channel estimates at the receiver, and thus a large loss in sum rate, whereas excessive training and feedback becomes too costly as a large portion of the frame is spent on overhead. 

\subsection{Training and Feedback Optimization}

Given the expression for system throughput with overhead, we propose solving the following optimization problem
\small
\begin{equation}
\max_{\tau_p,\tau_c} \left(\frac{T-(\tau_c+\tau_p)}{T}\right)\left(\mathbb{E}_{\bH}R_{sum}- \sum\limits_{i}d_i\log_2\left(1+\alpha \frac{\left(\frac{N_r^2}{\tau_p}+\frac{KN_tN_r}{\tau_c}\right)}{(KN_t-N_r)}\left(K-\frac{1}{d_i}\right)\right)\right),
\label{eqn:optimization_1}
\end{equation}
\normalsize
over the set of feasible training and feedback lengths. The sum rate expression defined in (\ref{eqn:sum_overhead}) is not convex as it is defined on a non-convex non-continuous closed set of bounded integers, as would be the case if one were optimizing over the number of feedback bits. Nevertheless, we seek to maximize a continuous relaxation of the defined cost function by typical methods of convex optimization~\cite{caire710multiuser, kobayashi2009training}.

\begin{figure}
\centering
\includegraphics[width=3.5in]{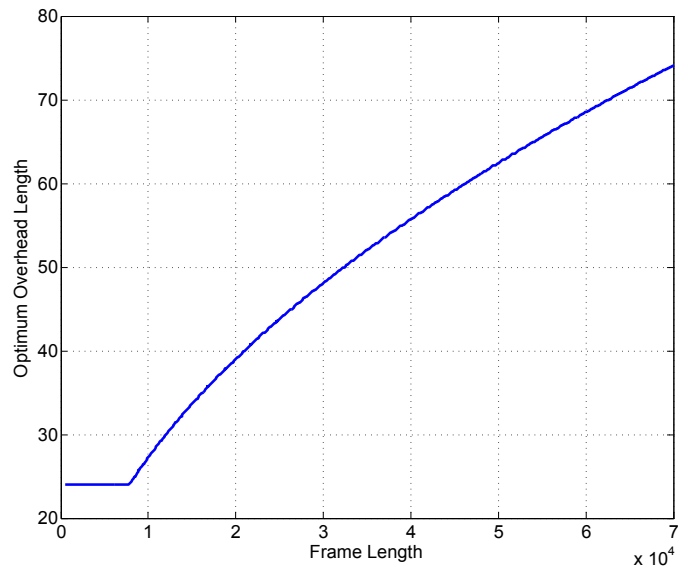}
\caption{Optimal Overhead Length vs. Frame Length for a 3 user $2 \times 2$ system with $d_i=1\ \forall i$ and $P_f=P$.}
\label{fig:T_vs_frame}
\end{figure}

The optimization can be simplified by realizing that for a fixed overhead length $T_{total}$, optimizing training and feedback lengths simplifies to minimizing $c_2(\tau_p,\tau_c)$ and then finding the optimal overhead length. So given a fixed amount of overhead the solution of the first optimization step can be shown to be
\begin{equation}
\tau_p=\frac{N_r}{N_r+\sqrt{KN_tN_r}}T_{total}, \quad \tau_c=\frac{\sqrt{KN_tN_r}}{N_r+\sqrt{KN_tN_r}}T_{total},
\label{eqn:time_allocation}
\end{equation}
which gives the optimal value $c_{opt}=\frac{(N_r+\sqrt{KN_tN_r})^2}{(KN_t-N_r)T_{total}}$. Replacing $c_{opt}$ into (\ref{eqn:optimization_1}), the optimal training length $T_{total}$ can now be found by solving
\small
\begin{align}
\begin{split}
\frac{\delta}{\delta T_{total}}\left(\frac{T-T_{total}}{T}\right)\left(\mathbb{E}_{\bH}R_{sum}-\sum\limits_{i}d_i\log_2\left(1+\frac{\alpha(K-1/d_i) \left(N_r+\sqrt{KN_tN_r}\right)^2}{(KN_t-N_r)T_{total}}\right)\right) & = 0 \\
-\frac{1}{T}\left(\mathbb{E}_{\bH}R_{sum}-\sum d_i\log\left(1+ \frac{\gamma_i}{T_{total}}\right)\right) + \frac{T-T_{total}}{T}\left(\sum \frac{d_i}{T_{total}(T_{total}+\gamma_i)}\right) & =0
\label{eqn:optimization}
\end{split}
\end{align}
\normalsize
where $\gamma_i=\frac{\alpha (K-1/d_i)(N_r+\sqrt{KN_tN_r})^2}{(KN_t-N_r)}$. Though the exact solution to this optimization cannot be found in closed form, it can be shown that $T_{total}$ increases with $\sqrt{T}$, and \emph{initially} decreases with the ratio transmit to feedback power $\frac{P}{P_f}$. Fig. \ref{fig:T_vs_frame} plots the optimal feedback vs. frame length, where we have solved the optimization in (\ref{eqn:optimization}) numerically for a three $2 \times 2$ user network with $d_i =1$. Fig. \ref{fig:T_vs_frame} verifies the claimed scaling and shows that for a range of reasonable frame lengths, the solution to the optimization problem is less than the minimum length required to satisfy the dimensionality constraints on the training and feedback matrices, i.e. the optimal overhead is minimal for realistic frame sizes.

\begin{figure}
\centering
\includegraphics[width=3.5in]{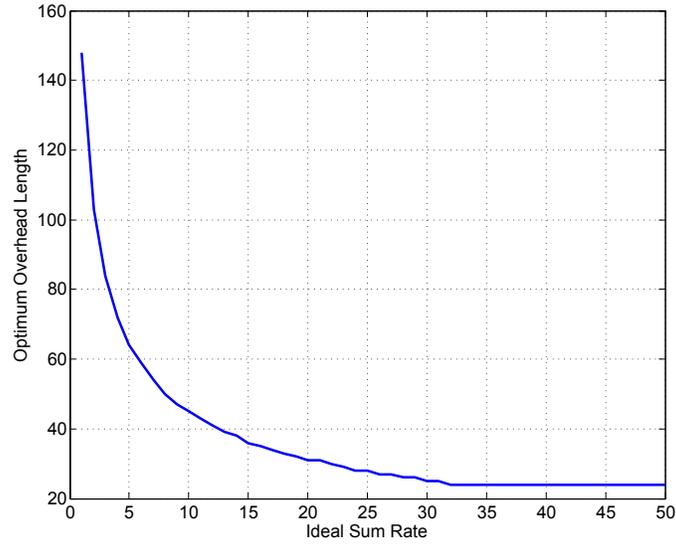}
\caption{Optimal Overhead length vs. $R_{sum}$ for a 3 user $2 \times 2$ system  with $d_i=1\ \forall i$, SNR=40dB and $P_f=P/10$.}
\label{fig:T_vs_R_sum}
\end{figure}

While it may not be surprising that longer frames can support more training~\cite{kobayashi2009training}, it's interesting to note that small mismatches in forward and feedback SNR initially decrease training and feedback lengths. Hence, a slightly noisy feedback channel does not require extra training to compensate. This, however, is not true of significantly poor feedback channels, as the optimal overhead length does in fact increase to improve the quality of CSI. It can also be shown that, all else fixed, the optimal training length decreases with the achieved sum rate or effectively SNR, making analog feedback especially efficient at high SNR. This is shown if Fig. \ref{fig:T_vs_R_sum}.

\section{Simulation Results} \label{sec:sims}

In this section we present simulation results that validate the claims and proofs given in Sections \ref{sec:mux_proof} and \ref{sec:overhead}. We verify the results shown in Theorems \ref{thm:sumratescaling} and \ref{thm:sumrate_slow_feedback} which state that as long as the transmit power on the feedback channel scales sufficiently with the power on the forward channel, the multiplexing gain achieved by perfect IA is preserved. To better show the performance of both IA and IA with analog feedback we remove the restriction of per-stream receivers and thus calculate the sum rate of a joint decoder which also treats interference as colored Gaussian noise  
\begin{equation}
R_{sum} = \sum_{i=1}^{K} \log_{2} \left| \mathbf{I} + \frac{P}{d_i} \mathbf{H}_{i,i}\mathbf{F}_{i}\mathbf{F}_{i}^*\mathbf{H}_{i,i}^*\left(\sigma^{2}\mathbf{I} + \sum_{k \neq i}\frac{P}{d_k}\bH_{i,k}\bF_k\bF_k^*\bH_{i,k}^*  \right)^{-1}\right|, \nonumber
\end{equation}
and the precoders are calculated given ideal or estimated CSI~\cite{Blum2003}.

\begin{figure}
\centering
\includegraphics[width=3.5in]{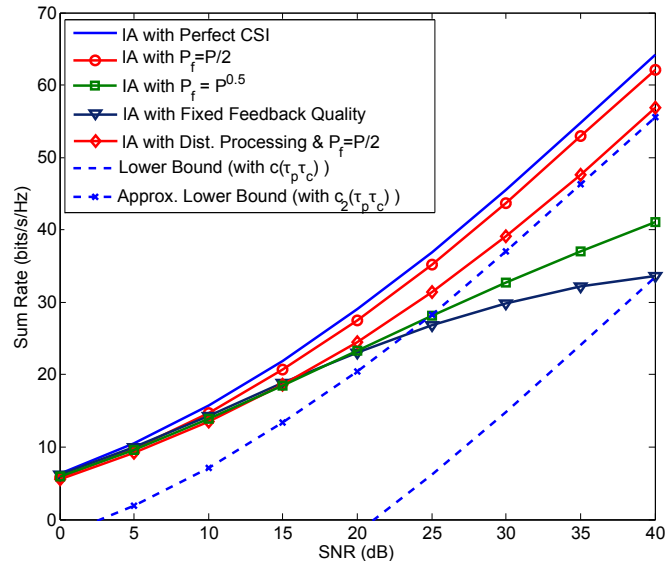}
\caption{The sum rate achieved by IA in a 3 users $5\times 4$ system ($d_i=2\ \forall i$) with perfect CSI, scaling feedback quality ($P_{f}=P/2$), slower scaling feedback ($P_{f}=\sqrt{P}$), and fixed quality feedback with $SNR_{f}=5dB$.}
\label{fig:sumrate}
\end{figure}

\begin{figure}
\centering
\includegraphics[width=3.5in]{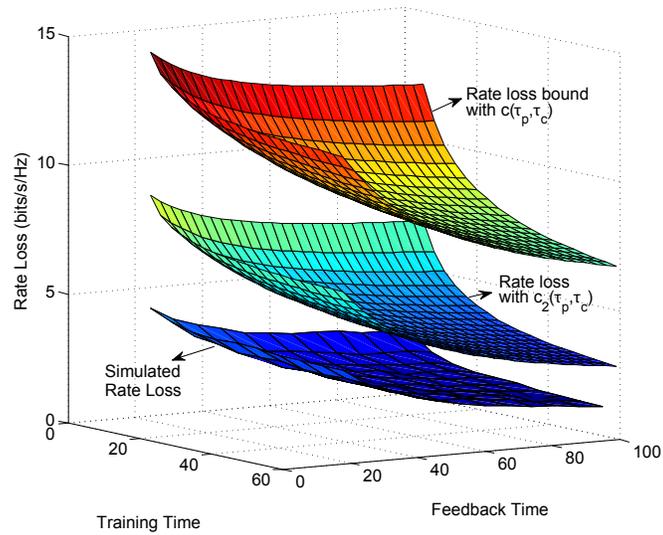}
	\caption{Rate loss vs. training and feedback times for a 3 user $2\times 2$ system with $d_i=1\ \forall i$ and $P_f=P/10$.}
\vspace{-22pt}
	\label{fig:loss_vs_time}
\end{figure}

Fig. \ref{fig:sumrate} shows the sum rate achieved by the IA algorithm in \cite{Peters2010} in a 3 user $5 \times 4$ system with $d_i=2\ \forall i$. We show performance with perfect CSI, scaling quality CSI where $P_{f}=P/2$, slower scaling CSI with $P_f=P^{\beta}$ and $\beta=0.5$ and fixed quality CSI where the SNR on the feedback channel is fixed at 5dB. In all cases, minimum training and feedback lengths are used, i.e. $\tau_p=12$ and $\tau_c=45$. Fig. \ref{fig:sumrate} confirms that both perfect and scaling feedback exhibit the same sum rate scaling or degrees of freedom.  This establishes the multiplexing gain optimality of using analog feedback. Fig. \ref{fig:sumrate} also confirms the fact that the mean loss in sum rate at high enough SNR is indeed a constant independent of the forward channel SNR. Moreover, Fig. \ref{fig:sumrate} shows that while the rate loss bound with $c(\tau_p,\tau_c)$ suffices to establish a constant rate loss and multiplexing gain optimality, it is loose for larger systems as mentioned in Section \ref{sec:mux_proof}. Using $c_2(\tau_p,\tau_c)$, by assuming a biunitary invariant error distribution, gives a better characterization of achieved sum rate. Moreover, both rate loss characterizations, as well as simulated rate loss, naturally decrease with $\tau_p$ and $\tau_c$ as stated in Section \ref{sec:mux_proof}; the trend can be seen in Fig. \ref{fig:loss_vs_time} for a $2\times 2$ system with $P_f=P/10$.
When perfect scaling feedback is not possible, Fig. \ref{fig:sumrate} shows the slower yet linear scaling in the case of $P_f=P^{\beta}$ which verifies the result shown in Theorem \ref{thm:sumrate_slow_feedback} on the preservation of a fraction of the system's original multiplexing gain. Finally, for the case of fixed feedback quality, multiplexing gain is zero and the sum rate saturates at high SNR. 

Fig. \ref{fig:sumrate} also shows the performance of the distributed processing approach introduced in Section \ref{sec:analog}. As stated earlier, the cooperation assumed in Section \ref{sec:analog} is only practical in certain cases such as cellular systems. If cooperation is not possible, either a central node can calculate the IA solution and feed it forward to the other sources, or nodes calculate precoders independently. The analysis of the centralized processor strategy is straightforward. It only adds Gaussian noise to the precoders due to feed forward; i.e. all nodes will use noisy versions of the same vector, $\widehat{\bff}_{i}^\ell+\widetilde{\bff}_{i}^\ell$ and $\widehat{\bw}_i^\ell+\widetilde{\bw}_{i}^\ell$. So, the interference terms are now given by 
\small
\begin{equation}
(\widehat{\bw}_i^\ell+\widetilde{\bw}_{i}^\ell)^*(\underbrace{\widehat{\bH}_{i,k}+\widetilde{\bH}_{i,k}}_{\bH_{i,k}})(\widehat{\bff}_{i}^\ell+\widetilde{\bff}_{i}^\ell)= (\widehat{\bw}_i^m)^*\underbrace{\widetilde{\bH}_{i,k}}_{P_f^{-1}}\widehat{\bff}_{k}^{\ell} +(\widehat{\bw}_i^\ell)^*\bH_{i,k}\underbrace{\widetilde{\bff}_{i}^\ell}_{P_f^{-1}} +\underbrace{(\widetilde{\bw}_i^\ell)^*}_{P_f^{-1}}\bH_{i,k}\widehat{\bff}_{i}^\ell +\underbrace{(\widetilde{\bw}_i^\ell)^*}_{P_f^{-1}}\bH_{i,k}\underbrace{\widetilde{\bff}_{i}^\ell}_{P_f^{-1}}. \nonumber
\label{eqn:frob_bound}
\end{equation}
\normalsize
All errors in this case again decay with SNR as indicated above, and similar bounds on rate loss can be readily derived. In the case of distributed processing, the extra loss is due to the mismatch of CSI between nodes; IA solutions based on different perturbations of the same channels will be mismatched leading to extra interference leakage. Fig. \ref{fig:sumrate}, however, shows that the extra loss due to distributed processing is small, and no degrees of freedom are lost. The performance of distributed processing is not theoretically surprising. Since IA solutions and algorithms depend heavily on invariant and singular subspaces, Wedin's theorem~\cite{wedin1972perturbation, stewart} and results in \cite{stewart1990perturbation} can be used to tractably bound the angles between singular and invariant subspaces of perturbed matrices. Using the bounds in~\cite{stewart} we can see that even if nodes compute precoders using different perturbed channels, the angle (or error) between the different precoders calculated is bounded by a linear function of $\|\widetilde{\bH}_{i,k}\|^2_F$. Therefore, errors still decay with feedback power. Such decay is all that is needed to prove the multiplexing gain preservation and a constant rate loss. Computing a tight constant bound on $\Delta R_{sum}$, however, becomes significantly more involved. The small loss due to distributed processing, and the fact that it does not need extra overhead, make it a practical and viable approach.

\begin{figure}
\centering
\includegraphics[width=3.5in]{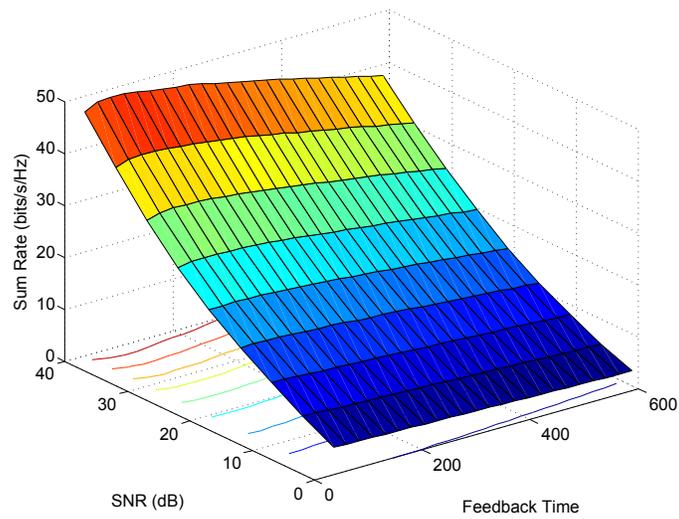}
\caption{The effective throughput achieved by interference alignment in a 3 user $5\times 4$ with $P_{f}=P/100$ and a frame length $T=2000$. This confirms that the amount of feedback needed to achieve optimal throughput is close to minimal.}
\label{fig:fb_var_snr}
\end{figure}

\begin{figure}
\centering
\includegraphics[width=3.5in]{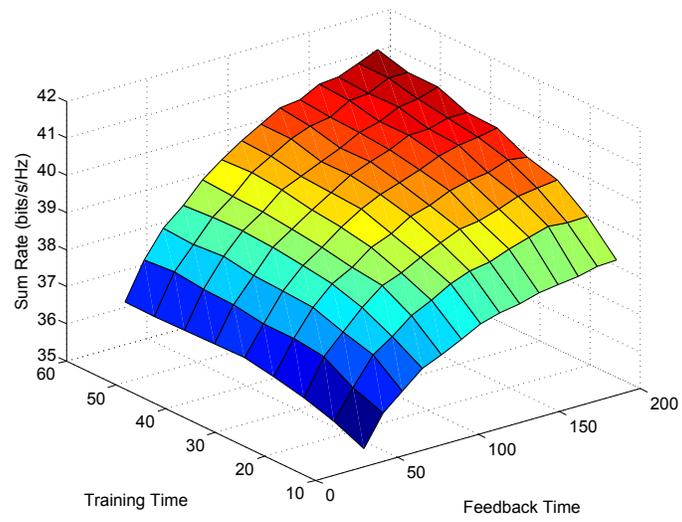}
\caption{This plots the sum rate achieved by IA in a 3 user $5\times 4$ at 35dB with and feedback SNR of 10dB and a frame length $T=10000$. 
}
\label{fig:training_vs_fb}
\end{figure}

Finally. we simulate the system's total throughput according to the overhead model presented in Section \ref{sec:overhead}. Fig. \ref{fig:fb_var_snr} shows that when the forward and reverse channel SNR scale together, the optimal feedback length is close to the theoretical minimum required for a frame length of 2,000, as predicted by Fig. \ref{fig:T_vs_frame} even for the relatively poor feedback channel ($P_f=P/100$) considered. Although analog feedback provides poor overhead scaling with frame length, the effect of this scaling is little since overhead is minimal for practical frames~\cite{kobayashi2009training}. Finally, Fig \ref{fig:fb_var_snr} shows the optimal feedback length assuming fixed training for ease of exposition. We know, however, from Section \ref{sec:overhead}, that this is not optimal. Fig. \ref{fig:training_vs_fb} shows that while generally less resources are used for training, both training and feedback must scale together for maximum throughput.
%


\section{Conclusions} \label{sec:Conclusion}

In this paper we proposed a low overhead feedback strategy for the interference channel. We showed that when combined with interference alignment, analog feedback can achieve full multiplexing gain when the forward and reverse channel SNR levels are comparable. When such symmetry is not possible, we showed a fraction of the degrees of freedom is retained. Thus a main benefit of analog feedback is that the cost of imperfect channel knowledge at the transmitter is bounded and quickly becomes negligible at high SNR. The mild requirement of comparable feedback and transmit power implies that analog feedback performs well with constant overhead, in the high SNR regime where IA is optimal. In addition to quantifying the cost of imperfect CSI, we show the scaling of required overhead with several network variables such as SNR and frame length. In simulation, we show that the throughput loss due to the overhead of training and analog feedback is often minimal. 
\vspace{-5pt}


\appendices 
\section{Proof of Lemma \ref{lemma:identical}} \label{sec:proof_lemma}
Consider \cite{Cadambe2008, Gomadam2008, Peters2010} or similar solutions to finding unitary IA precoders $\bF_i=\left[\bff_i^1, \hdots, \bff_i^{d_i}\right]$. In all such solutions, $\bF_i\ \forall i$ are functions of all interfering channels $\bH_{k,\ell}\ \forall k,\ell,\  k \neq \ell$ only. Since $\bF_i$ is not a function of $\bH_{i,i}$, and channels are independent across users, this implies that $\bF_i$ is independent of $\bH_{i,i}\ \forall i$. Therefore, $\bH_{i,i}\bff_i^m$ are Gaussian vectors with covariance $\mathbb{E}\left[\bH_{i,i}\bff_i^m(\bff_i^m)^*\bH_{i,i}^*\right]=\mathrm{trace}\left(\bff_i^m(\bff_i^m)^*\right)\bI_{N_r}=\bI_{N_r}$ \cite[Ch. 21]{mathai1992quadratic}.  
Similarly, due to the unitary precoders, $\mathbb{E}\left[\bH_{i,i}\bff_i^m(\bff_i^\ell)^*\bH_{i,i}^*\right]= \mathrm{trace}\left(\bff_i^m(\bff_i^\ell)^*\right)\bI_{N_r}=\mathbf{0}_{N_r}$, and therefore $\bH_{i,i}\bff_i^m$ and $\bH_{i,i}\bff_i^\ell$ are independent $\forall \ell \neq m$. Vectors $\widehat{\bff}_i^m$ are calculated similarly based on $\widehat{\bH}_{k,\ell}\ \forall k,\ell,\  k \neq \ell$ and are thus independent of both $\widehat{\bH}_{i,i}$ and $\bH_{i,i}$, thus $\bH_{i,i}\widehat{\bff}_i^\ell$ satisfy the same properties as $\bH_{i,i}\bff_i^\ell$. 

The combiners $\bw_i^m$ must now satisfy (\ref{eqn:conditions1}); for continuously distributed i.i.d channels (\ref{eqn:conditions2}) will be satisfied automatically~\cite{Cadambe2008}. This can be done by letting
\begin{equation}
\bw_i^m=\bU_{min}\left(\left[\bH_{i,1}\bF_1, \hdots, \bH_{i,i-1}\bF_{i-1}, \bH_{i,i}\bF_i^{(m)},\bH_{i,i+1}\bF_{i+1},\hdots, \bH_{i,K}\bF_K \right]\right),
\label{eqn:w}
\end{equation}
where $\bF_i^{(m)}=\left[\bff_1^1,\hdots\bff_1^{m-1},\bff_1^{m+1}, \hdots,\bff_1^{d_i}\right]$, and $\bU_{min}(\bA)$ extracts the least dominant left singular vector of $\bA$. Given that IA is feasible, the interference matrix in \eqref{eqn:w} will span at most $N_r-1$ dimensions (at most $N_r-d_i$ and $d_i-1$ dimensions of inter-user and inter-stream interference respectively). Hence, $\bw_i^m$ will always correspond to a 0 singular value, thus satisfying (\ref{eqn:conditions1}). 

So $\bw_i^m$ are a function of vectors $\bH_{i,k}\bff_k^\ell\ \forall k\neq i, \forall \ell$ and $\bH_{i,i}\bff_i^\ell,\ \forall \ell\neq m$. But $\bH_{i,k}\bff_k^\ell\ \forall k\neq i, \forall \ell$ are independent of $\bH_{i,i}\bff_i^m$ due to the independence of channels, and $\bH_{i,i}\bff_i^\ell,\ \forall \ell\neq m$ are independent of $\bH_{i,i}\bff_i^m$ as shown earlier. Thus $\bw_i^m$ is independent of $\bH_{i,i}\bff_i^m$. Since $\bw_i^m$ and $\bH_{i,i}\bff_i^m$ are independent, and since $\bH_{i,i}\bff_i^m$ has a unitary invariant Gaussian distribution as shown earlier \cite{tulino2004random}, we can perform a change of basis such that $\bw_i^m=\left[1, 0, \hdots, 0\right]^*$. After this change of basis $(\bw_i^m)^*\bH_{i,i}\bff_i^m$ is simply the first element of $\bH_{i,i}\bff_i^m$ which we have shown above is $\mathcal{CN}(0,1)$. Therefore, the terms $\left|(\bw_i^m)^*\bH_{i,i}\bff_i^m\right|^2$ are exponentially distributed with parameter 1. 

The vectors $\widehat{\bw}_i^m$ are again given by \eqref{eqn:w} using $\widehat{\bH}_{k,\ell}$ and $\widehat{\bff}_k^\ell$ instead of $\bH_{k,\ell}$ and $\bff_k^\ell$ respectively, and are thus again functions of variables independent of both $\widehat{\bH}_{i,i}\widehat{\bff}_i^m$ and $\bH_{i,i}\widehat{\bff}_i^m$. Thus $\widehat{\bw}_i^m$ are independent of $\bH_{i,i}\widehat{\bff}_i^m$ and again a change of basis reveals that $(\widehat{\bw}_i^m)^*\bH_{i,i}\widehat{\bff}_i^m$ are $\mathcal{CN}(0,1)$ and $\left|(\widehat{\bw}_i^m)^*\bH_{i,i}\widehat{\bff}_i^m\right|^2$ are exponentially distributed with parameter 1. Therefore, $\left|(\bw_i^m)^*\bH_{i,i}\bff_i^m\right|^2$ and $\left|(\widehat{\bw}_i^m)^*\bH_{i,i}\widehat{\bff}_i^m\right|^2$ are identically and exponentially distributed with parameter 1.


\singlespace
\bibliographystyle{IEEEtran}
\bibliography{IEEEabrv,analog_IA}

\begin{thebibliography}{10}
\providecommand{\url}[1]{#1}
\csname url@samestyle\endcsname
\providecommand{\newblock}{\relax}
\providecommand{\bibinfo}[2]{#2}
\providecommand{\BIBentrySTDinterwordspacing}{\spaceskip=0pt\relax}
\providecommand{\BIBentryALTinterwordstretchfactor}{4}
\providecommand{\BIBentryALTinterwordspacing}{\spaceskip=\fontdimen2\font plus
\BIBentryALTinterwordstretchfactor\fontdimen3\font minus
  \fontdimen4\font\relax}
\providecommand{\BIBforeignlanguage}[2]{{%
\expandafter\ifx\csname l@#1\endcsname\relax
\typeout{** WARNING: IEEEtran.bst: No hyphenation pattern has been}%
\typeout{** loaded for the language `#1'. Using the pattern for}%
\typeout{** the default language instead.}%
\else
\language=\csname l@#1\endcsname
\fi
#2}}
\providecommand{\BIBdecl}{\relax}
\BIBdecl

\bibitem{Cadambe2008}
V.~Cadambe and S.~Jafar, ``Interference alignment and degrees of freedom of the
  {K}-user interference channel,'' \emph{IEEE Transactions on Information
  Theory}, vol.~54, no.~8, pp. 3425 --3441, Aug. 2008.

\bibitem{GuoJafar}
T.~Gou and S.~Jafar, ``Degrees of freedom of the {K} user {MxN} {MIMO}
  interference channel,'' \emph{IEEE Transactions on Information Theory},
  vol.~56, no.~12, pp. 6040 --6057, Dec. 2010.

\bibitem{Peters2010}
S.~Peters and R.~W. Heath, Jr., ``Cooperative algorithms for {MIMO}
  interference channels,'' \emph{IEEE Transactions on Vehicular Technology},
  vol.~60, no.~1, pp. 206 --218, Jan. 2011.

\bibitem{Yetis2009}
C.~Yetis, T.~Gou, S.~Jafar, and A.~Kayran, ``On feasibility of interference
  alignment in {MIMO} interference networks,'' \emph{IEEE Transactions on
  Signal Processing}, vol.~58, no.~9, pp. 4771 --4782, Sept. 2010.

\bibitem{Ayach2009}
{O. El Ayach}, S.~Peters, and R.~W. Heath, Jr., ``The feasibility of
  interference alignment over measured {MIMO-OFDM} channels,'' \emph{IEEE
  Transactions on Vehicular Technology}, vol.~59, no.~9, pp. 4309 --4321, Nov.
  2010.

\bibitem{Gomadam2008}
K.~Gomadam, V.~Cadambe, and S.~Jafar, ``{Approaching the capacity of wireless
  networks through distributed interference alignment},'' \emph{Proc. of IEEE
  Global Telecommunications Conference}, pp. 1 --6, Nov. 30 - Dec. 4 2008.

\bibitem{Berry-BidirectionalIA}
\BIBentryALTinterwordspacing
C.~{Shi}, R.~A. {Berry}, and M.~L. {Honig}, ``{Adaptive Beamforming in
  Interference Networks via Bi-Directional Training},'' \emph{ArXiv preprint
  arXiv:1003.4764}, Mar. 2010. [Online]. Available:
  \url{http://arxiv.org/abs/1003.4764}
\BIBentrySTDinterwordspacing

\bibitem{MMSE-IA}
D.~Schmidt, C.~Shi, R.~Berry, M.~Honig, and W.~Utschick, ``Minimum mean squared
  error interference alignment,'' \emph{Proc. of the Forty-Third Asilomar
  Conference on Signals, Systems and Computers}, pp. 1106 --1110, Nov. 2009.

\bibitem{Choi2009}
S.~W. Choi, S.~Jafar, and S.-Y. Chung, ``On the beamforming design for
  efficient interference alignment,'' \emph{IEEE Communications Letters},
  vol.~13, no.~11, pp. 847 --849, Nov. 2009.

\bibitem{Tresch}
R.~Tresch, M.~Guillaud, and E.~Riegler, ``On the achievability of interference
  alignment in the {K}-user constant {MIMO} interference channel,'' \emph{Proc.
  IEEE/SP 15th Workshop on Statistical Signal Processing}, pp. 277--280, Aug.
  31-Sept. 3 2009.

\bibitem{santamaria-max-sum-rate}
I.~Santamar{\'\i}a, O.~Gonz{\'a}lez, R.~W. Heath, Jr., and S.~W. Peters,
  ``Maximum sum-rate interference alignment algorithms for {MIMO} channels,''
  \emph{Proc. of IEEE Global Telecommunications Conference}, Dec. 2010.

\bibitem{dimakis}
\BIBentryALTinterwordspacing
D.~S. Papailiopoulos and A.~G. Dimakis, ``{Interference alignment as a rank
  constrained rank minimization},'' \emph{Arxiv preprint arXiv:1010.0476},
  2010. [Online]. Available: \url{http://arxiv.org/abs/1010.0476}
\BIBentrySTDinterwordspacing

\bibitem{mimo_hardware_recip}
V.~Jungnickel, U.~Kruger, G.~Istoc, T.~Haustein, and C.~von Helmolt, ``A {MIMO}
  system with reciprocal transceivers for the time-division duplex mode,''
  \emph{Proc. of IEEE Antennas and Propagation Society International
  Symposium}, vol.~2, pp. 1267 -- 1270, Jun. 2004.

\bibitem{guillaud_recip}
M.~Guillaud, D.~Slock, and R.~Knopp, ``A practical method for wireless channel
  reciprocity exploitation through relative calibration,'' \emph{Proc. of the
  Eighth International Symposium on Signal Processing and Its Applications},
  pp. 403 -- 406, Aug. 2005.

\bibitem{Peters2010a}
S.~Peters and R.~W. Heath, Jr., ``{Orthogonalization to reduce overhead in MIMO
  interference channels},'' \emph{Proc. of International Zurich Seminar}, pp.
  126 --129, Mar. 2010.

\bibitem{Thukral2009}
H.~Bolcskei and {J. Thukral}, ``Interference alignment with limited feedback,''
  \emph{Proc. IEEE International Symposium on Information Theory}, pp. 1759
  --1763, Jun. 28 -Jul. 3 2009.

\bibitem{Krishnamachari2009}
\BIBentryALTinterwordspacing
R.~Krishnamachari and M.~Varanasi, ``{Interference Alignment Under Limited
  Feedback for MIMO Interference Channels},'' \emph{Arxiv preprint
  arXiv:0911.5509}, 2009. [Online]. Available:
  \url{http://arxiv.org/abs/0911.5509}
\BIBentrySTDinterwordspacing

\bibitem{Jindal2006}
N.~Jindal, ``{MIMO} broadcast channels with finite-rate feedback,'' \emph{IEEE
  Transactions on Information Theory}, vol.~52, no.~11, pp. 5045 --5060, Nov.
  2006.

\bibitem{Marzetta2006}
T.~Marzetta and B.~Hochwald, ``{Fast transfer of channel state information in
  wireless systems},'' \emph{IEEE Transactions on Signal Processing}, vol.~54,
  no.~4, pp. 1268--1278, Apr. 2006.

\bibitem{Samardzija2006}
D.~Samardzija and N.~Mandayam, ``{Unquantized and uncoded channel state
  information feedback in multiple-antenna multiuser systems},'' \emph{IEEE
  Transactions on Communications}, vol.~54, no.~7, pp. 1335--1345, Jul. 2006.

\bibitem{caire710multiuser}
G.~Caire, N.~Jindal, M.~Kobayashi, and N.~Ravindran, ``Multiuser {MIMO}
  achievable rates with downlink training and channel state feedback,''
  \emph{IEEE Transactions on Information Theory}, vol.~56, no.~6, pp. 2845
  --2866, Jun. 2010.

\bibitem{shanechi-comparison-of-practical-feedback}
M.~M. Shanechi, R.~Porat, and U.~Erez, ``Comparison of practical feedback
  algorithms for multiuser {MIMO},'' \emph{IEEE Transactions on
  Communications}, vol.~58, no.~8, pp. 2436 --2446, Aug. 2010.

\bibitem{Ayach2010}
{O. El Ayach} and R.~W. Heath, Jr., ``Interference alignment with analog {CSI}
  feedback,'' \emph{Proc. of IEEE Military Communications Conference}, pp. 1644
  --1648, Oct. 31-Nov. 3 2010.

\bibitem{martinian-waterfilling}
E.~Martinian, ``{Waterfilling gains O (1/SNR) at high SNR},'' \emph{Unpublished
  notes available from http://www.csua.berkeley. edu/\~{}
  emin/research/wfill.pdf.}

\bibitem{Rose_Greedy}
C.~Rose, S.~Ulukus, and R.~Yates, ``Wireless systems and interference
  avoidance,'' \emph{IEEE Transactions on Wireless Communications}, vol.~1,
  no.~3, pp. 415 --428, Jul. 2002.

\bibitem{love-heath-limited-feedback-unitary}
D.~Love and R.~W. Heath, Jr., ``Limited feedback unitary precoding for spatial
  multiplexing systems,'' \emph{IEEE Transactions on Information Theory},
  vol.~51, no.~8, pp. 2967 --2976, Aug. 2005.

\bibitem{makouei-simple-sinr-characterization}
B.~Nosrat-Makouei, J.~Andrews, and R.~W. Heath, Jr., ``A simple {SINR}
  characterization for linear interference alignment over uncertain {MIMO}
  channels,'' \emph{Proc. of IEEE Int. Symposium on Inf. Theory}, pp. 2288
  --2292, Jun. 2010.

\bibitem{Caire_shamai}
G.~Caire, N.~Jindal, and S.~Shamai, ``On the required accuracy of transmitter
  channel state information in multiple antenna broadcast channels,''
  \emph{Proc. of the Forty-First Asilomar Conference on Signals, Systems and
  Computers}, pp. 287 --291, Nov. 2007.

\bibitem{marzetta1999blast}
T.~Marzetta, ``{BLAST Training: Estimating Channel Characteristics for
  High-Capacity Space-Time Wireless},'' \emph{Proc. of 37th Annual Allerton
  Conference on Communications, Control, and Computing}, Sept. 1999.

\bibitem{chae2008coordinated}
C.~Chae, D.~Mazzarese, T.~Inoue, and R.~W. Heath, Jr., ``{Coordinated
  beamforming for the multiuser MIMO broadcast channel with limited
  feedforward},'' \emph{IEEE Trans. Signal Process.}, vol.~56, no.~12, pp.
  6044--6056, Dec. 2008.

\bibitem{Blum2003}
R.~Blum, ``{MIMO capacity with interference},'' \emph{IEEE J. Sel. Areas
  Commun.}, vol.~21, no.~5, pp. 793--801, Jun. 2003.

\bibitem{honigblind}
M.~Honig, U.~Madhow, and S.~Verdu, ``Blind adaptive multiuser detection,''
  \emph{IEEE Transactions on Information Theory}, vol.~41, no.~4, pp. 944
  --960, Jul. 1995.

\bibitem{kobayashi2009training}
\BIBentryALTinterwordspacing
M.~Kobayashi, N.~Jindal, and G.~Caire, ``{Training and Feedback Optimization
  for Multiuser MIMO Downlink},'' \emph{to appear in IEEE Transactions on
  Communications}, 2011. [Online]. Available:
  \url{http://arxiv.org/abs/0912.1987}
\BIBentrySTDinterwordspacing

\bibitem{wedin1972perturbation}
P.~Wedin, ``{Perturbation bounds in connection with singular value
  decomposition},'' \emph{BIT Numerical Mathematics}, vol.~12, no.~1, pp.
  99--111, 1972.

\bibitem{stewart}
G.~Stewart, ``Perturbation theory for the singular value deomposition,''
  \emph{SVD and Signal Processing II, Algorithms, Analysis and Applications
  (Elsevier Science)}, pp. 99 --109, 1991.

\bibitem{stewart1990perturbation}
------, ``{Stochastic perturbation theory},'' \emph{SIAM review}, vol.~32,
  no.~4, pp. 579--610, 1990.

\bibitem{mathai1992quadratic}
A.~Mathai and S.~Provost, \emph{{Quadratic forms in random variables: theory
  and applications}}.\hskip 1em plus 0.5em minus 0.4em\relax M. Dekker, New
  York, 1992.

\bibitem{tulino2004random}
A.~Tulino and S.~Verd{\'u}, \emph{{Random matrix theory and wireless
  communications}}.\hskip 1em plus 0.5em minus 0.4em\relax Now Publishers Inc,
  2004.

\end{thebibliography}

\newpage

\end{document}